# Temporal Weyl Physics and Topological Control of Direction-Selected Radiation in Anisotropic Photonic Time Crystals

Zihao He[1], Sihao Zhang[2], Huanan Li[2†] and Xiang Ni[1‡]

[1]School of Physics, Central South University, Changsha, Hunan 410083, China

[2]MOE Key Laboratory of Weak-Light Nonlinear Photonics, School of Physics, Nankai University, Tianjin 300071, China

†Contact author: hli01@nankai.edu.cn
‡Contact author: nixiang@csu.edu.cn

*Anisotropic photonic time crystals (APTCs), enabled by periodic temporal modulation of a uniform anisotropic medium, exhibit asymmetric momentum-bandgap structures and offer unique control over light–matter interactions. Here, we introduce and construct temporal Weyl points (TWPs) in APTCs within a synthetic three-dimensional space defined by two phase parameters and the quasi-frequency. The temporal response reveals robust Fermi arcs linking TWPs of opposite topological charge. Unlike spatial counterparts, these Fermi arcs emerge only after the first temporal supercell comprising multiple periods of APTCs, reflecting causality. We further show that TWPs generate a directional near-zero radiation trajectory in momentum space with tunable radiation from stationary charges embedded in APTCs, while the associated Fermi arcs robustly suppress radiation at selected directions and frequencies. Our findings establish temporal Weyl physics in photonic time crystals and uncover new opportunities for topological control of light–matter interactions through the time dimension.*

**Introduction**

Photonic time crystals (PTCs), created through periodic temporal modulation of a uniform medium, have opened new avenues for controlling light-matter interactions[1-3]. When electromagnetic waves encounter the time interfaces of PTCs, they undergo reflection and refraction in time, producing waves that propagate in oppose spatial directions with different frequencies[4-6]. The interference between these reflected and refracted waves in PTCs forms momentum bands and bandgaps that are fundamentally distinctive from their spatial counterparts[7-11]. Unlike spatial photonic crystals, where energy conservation holds, the breaking of time-translational symmetry in PTCs invalidates energy conservation[12]. Correspondingly, their momentum bandgaps if exist, can support exponentially growing modes that draw energy from modulation itself. Recent experiments have confirmed time refraction, time reflection, and the emergence of momentum bandgaps[13,14]. These exotic properties of PTC enable broadband amplification and absorption[15], temporal Faraday rotation[16,17], momentum bandgaps expansion[18,19], free-electron radiation[20], spontaneous emission[21] and topological temporal localized states[22-25]. Very recently, anisotropic photonic time crystals (APTCs) have emerged as a new class of PTCs, in which the relative permittivity tensor $\bar{\bar{\varepsilon}}(t)$ of a uniform anisotropic medium undergoes periodic temporal modulation. APTCs have attracted significant research interest, and several unique phenomena have been theoretically predicted, including stationary charge radiation[26], nonuniform wave momentum bandgap[11] and higher-order exceptional points[27].

On another research frontier, Weyl semimetals have attracted tremendous attention due to their exceptional properties, such as Fermi-arc surface states and chiral anomalies, which originate from topologically protected Weyl points in momentum space[28,29]. Traditionally, Weyl points are conceptualized as magnetic monopoles in the three-dimensional (3D) reciprocal space of crystalline materials. These topologically protected Weyl points guarantee the existence of topological surface states, which form Fermi arcs in the surface Brillouin zone, connecting Weyl points of opposite charge. Fermi arcs possess an open Fermi surface[30,31], a characteristic that is fundamentally distinct from conventional closed Fermi surfaces and leads to intriguing phenomena, like manipulating dispersion relations[32,33] and Andreev reflection[34]. Nevertheless, realizing Weyl semimetals in electronic systems remain challenging, with only limited experimental demonstrations[35-37]. In contrast, Weyl points and Fermi arcs have been extensively explored in artificial materials like photonics[37-41], acoustics[42,43], plasmonics[44], and circuit systems[45], where they have unveiled a wealth of topological wave phenomena. Recent advances in synthetic dimensions have enabled the realization of higher-dimensional topological phenomena like Weyl points and Fermi arcs within low-dimensional systems, substantially reducing experimental complexity[46-48]. This raises a compelling open question: can Weyl points and their associated Fermi arcs be engineered in synthetic higher dimensions within low dimensional APTCs? Furthermore, if Weyl points and Fermi arcs do exist, how might they influence, or even control the radiation of stationary charge embedded in APTCs?

In this work, we construct temporal Weyl points (TWPs) in the synthetic reciprocal space through a judicious design of APTCs, where the synthetic dimension is formed by two phase parameters together with a quasi-frequency. We demonstrate that these TWPs are protected by nonzero topological charge via rigorous Wilson loop calculation[47]. By introducing temporal cladding slabs (TCSs) to create temporal boundaries[24,49], Fermi arc will emerge in the synthetic reciprocal space, connecting TWPs with opposite topological charges. These Fermi arc modes localize at the temporal boundaries and persist robustly regardless of variations in the temporal boundary, serving as the surface states of a synthetic 3D structure. Unlike Fermi arcs in spatial Weyl semimetals, temporal Fermi arcs can only emerge after the first temporal supercell comprising several periods of APTCs and one TCS, a constraint imposed by causality. Notably, spontaneous radiation phenomena have been found in APTCs embedding with stationary charges[26]. Our results reveal two distinct influences of TWPs and Fermi arcs on the radiation behavior of stationary charge: TWPs generate a directional near-zero radiation trajectory compared to the radiation from momentum bandgap, and the associated Fermi arcs robustly suppress radiation at selected directions and frequencies.

**Frequency vs. momentum eigenvalue equations**

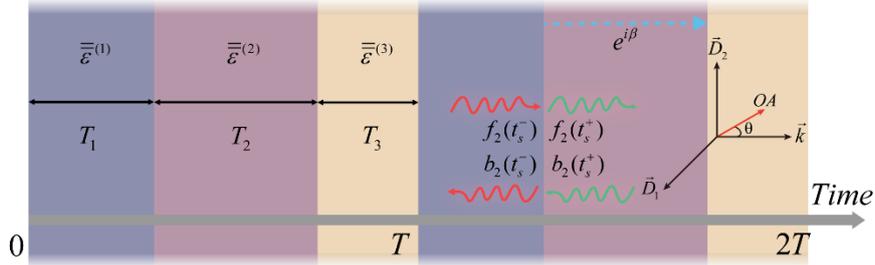

FIG. 1 An anisotropic photonic time crystal (APTC) comprising three temporally switched anisotropic media. The modulation time of each temporal slab $T_m (m=1,2,3)$ is determined by its extraordinary wave refractive $n_2^{(m)}$ of polar coordinate $\theta$, which is the angle between the wavevector $\vec{k}$ and the fixed optical axis (OA). The momentum phase accumulated in each slab $\beta$ is chosen to be equal for a given direction.

The investigated APTC is composed of $N=3$ layers of lossless, uniform, nonmagnetic uniaxial crystals, switched in time, with a dielectric tensor varying periodically in time $\bar{\bar{\varepsilon}}(t) = \bar{\bar{\varepsilon}}(t+T)$, where $T$ is the modulation period (see FIG. 1). We prove that Weyl-point degeneracies cannot occur when the number of layers in the temporal unit cell of an APTC satisfies $N \neq 3, N \in ¥$ [50]. Without loss of generality, we adopt the principal axes of the uniaxial crystal as the coordinate frame, fixing the optical axis (OA) along the $z$-direction. Consequently, the dielectric tensor for each medium $m$ ($m = 1,2,3$) is described by $\bar{\bar{\varepsilon}}^{(m)} = \mathrm{diag}\{\varepsilon_\perp^{(m)}, \varepsilon_\perp^{(m)}, \varepsilon_P^{(m)}\}$, with its components following a sinusoidal temporal modulation:

$$\varepsilon_\perp^{(m)} = \varepsilon_\perp + \delta\varepsilon_\perp \sin(\frac{2m\pi}{3} + \theta_1), \ \varepsilon_P^{(m)} = \varepsilon_P + \delta\varepsilon_P \sin(\frac{2m\pi}{3} + \theta_2), \qquad (1)$$

where $\varepsilon_\perp$ and $\varepsilon_P$ are modulation constants, $\delta\varepsilon_\perp$ and $\delta\varepsilon_P$ represent the modulation amplitudes, and $\theta_1$, $\theta_2 \in [0, 2\pi)$ are phase parameters. For a given wave vector $\vec{k} = [k_x, k_y, k_z]^\mathbf{T}$ (**T** denoting the transpose operation), the dispersion relation of a uniaxial crystal supports two distinct frequencies, $\omega_1^{(m)}$ and $\omega_2^{(m)}$, corresponding to ordinary and extraordinary waves. There are given by $\omega_1^{(m)} = kc_0 \big/ \sqrt{\varepsilon_\perp^{(m)}}$ and

$\omega_2^{(m)} = kc_0\sqrt{\hat{k}_x^2 + \hat{k}_y^2 / \varepsilon_P^{(m)} + \hat{k}_z^2 / \varepsilon_\perp^{(m)}}$ , where $c_0$ is the speed of light in a vacuum and $\hat{k}_\alpha = k_\alpha / |\vec{k}|$ with $\alpha = x, y, z$ [26]. The corresponding electric displacement vectors, $\vec{D}_1(\vec{k}) = [-\hat{k}_y, \hat{k}_x, 0]^T$ and $\vec{D}_2(\vec{k}) = [\hat{k}_x\hat{k}_z, \hat{k}_y\hat{k}_z, -(\hat{k}_x^2 + \hat{k}_y^2)]^T$, are mutually orthogonal and form an orthogonal triplet together with $\vec{k}$. Given a fixed OA, momentum conservation ensures that extraordinary and ordinary waves evolve independently in the APTC[50]. This unique feature allows us to exclusively investigate the properties of extraordinary wave. The momentum-space representations of the electric displacement field $\vec{D}_2(\vec{k})$ in temporal slab $m$ can be expressed as

$$\vec{D}_2^{(m)}(\vec{k},t) = \vec{D}_2(\vec{k})[1, 1]\psi_2(t), \tag{2}$$

where the vector $\psi_2(t) \equiv [f_2(t), b_2(t)]^T$ consists of the forward (and backward) time-dependent components $f_2(t) \propto e^{j\omega_2^{(m)}t}$ [and $b_2(t) \propto e^{-j\omega_2^{(m)}t}$]. The modulation period $T_m$ on each temporal slabs are set proportional to $n_2^{(m)} = 1/\sqrt{\sin^2\theta/\varepsilon_P^{(m)} + \cos^2\theta/\varepsilon_\perp^{(m)}}$, which is the extraordinary-wave refractive index in a specific polar coordinate $\theta$. This configuration ensures that the momentum phase (propagation phase) $\beta = \omega_2^{(m)}T_m$ on each temporal slabs is independent of index $m$ in a given direction (i.e., we chose $\theta = \pi/6$ throughout the text)[23,24]. Based on temporal boundary conditions and Floquet's theorem $D_2^{(m+3)} = e^{i\varphi}D_2^{(m)}$ [where $D_2^{(m)}$ represents the field right after the time interface between temporal slab $m$ and slab $m-1$, $\varphi = \Omega T \in (0, 2\pi]$ is the Floquet phase and $\Omega$ is the quasi-frequency], we define a new state vector $|u\rangle = [D_2^{(1)}, D_2^{(2)}, D_2^{(3)},]^T$ and derive the momentum eigenvalue equation as[50]

$$H_{eff}(\theta_1, \theta_2, \varphi)|u\rangle = \cos[\beta(\theta_1, \theta_2, \varphi)]|u\rangle, \tag{3}$$

with

$$H_{eff}(\theta_1, \theta_2, \varphi) = \begin{pmatrix} 0 & \frac{n_2^{(1)}}{n_2^{(2)} + n_2^{(1)}} & \frac{n_2^{(2)}}{n_2^{(2)} + n_2^{(1)}} \cdot e^{-i\varphi} \\ \frac{n_2^{(3)}}{n_2^{(3)} + n_2^{(2)}} & 0 & \frac{n_2^{(2)}}{n_2^{(3)} + n_2^{(2)}} \\ \frac{n_2^{(3)}}{n_2^{(3)} + n_2^{(1)}} \cdot e^{i\varphi} & \frac{n_2^{(1)}}{n_2^{(3)} + n_2^{(1)}} & 0 \end{pmatrix}. \tag{4}$$

Two phase parameters $\theta_1$ and $\theta_2$, together with $\varphi$, constitute a synthetic reciprocal space ($\theta_1, \theta_2, \varphi$) that might enable the emergence of topological wave phenomena in temporally modulated system. Eq. (3) facilitates the calculation of momentum bands $\beta_n$ ($n = 1, 2, 3$) of APTCs in the synthetic reciprocal space and corresponding eigenstates $|u_n\rangle$, allows for the subsequent evaluation of their topological properties. While our investigation focuses on periodic APTCs, the approach above can naturally extend to the study of photonic time quasicrystals and defective structures[50].

**Synthetic Weyl points in APTCs**

We consider the range of $\beta \in (0, \pi]$ for simplicity due to the even symmetry of eigenvalue in Eq. (3). By scanning the entire synthetic reciprocal space, we identify twofold-degenerate points at momentum phase $\beta = \pi/3$ in the $\varphi = \pi$ plane, and the two bands intersect linearly, which we refer to as TWPs [see FIG.2 (a)]. From Eq. (3), the characteristic equation governing the eigenvalues $\lambda_n = \cos\beta_n$ can be derived as

$$\lambda^3 + (2X - 1)\lambda + 2\cos\varphi X = 0, \tag{5}$$

where $X = n_2^{(1)}n_2^{(2)}n_2^{(3)} / [(n_2^{(1)} + n_2^{(2)})(n_2^{(2)} + n_2^{(3)})(n_2^{(3)} + n_2^{(1)})]$. When the discriminant of Eq. (5) meets $\Delta = 0$, two identical solutions $\lambda$ are obtained (see ref. [50] for details), such that

$$\Delta = \cos^2\varphi X^2 + \left(\frac{2X-1}{3}\right)^3 = 0. \qquad (6)$$

Since all $n_2^{(m)}$ are positive and real, $X$ is limited to the range $(0, 1/8]$, which leads to the discriminant $\Delta \leq 0$. Note that $\Delta$ attains its maximum value of 0 only when both conditions are simultaneously satisfied: $\cos^2\varphi = 1$ and $X = 1/8$ [50]. The blue, red, and green dashed lines in FIG. 2(b) indicate the condition $n_2^{(1)} = n_2^{(2)}$, $n_2^{(1)} = n_2^{(3)}$, and $n_2^{(2)} = n_2^{(3)}$ in the synthetic reciprocal space, respectively. Therefore, the intersections of these lines, implying the condition $n_2^{(1)} = n_2^{(2)} = n_2^{(3)}$, precisely correspond to the TWPs' locations as $X$ only attains the value 1/8 in this case. For the special case where $\varphi = \pi$ ($\varphi = 2\pi$), a trivial solution $\lambda_1 = \cos\beta_1 = -1$ (+1) in Eq. (5) always exists, leading to the emergence of a trivial momentum band $\beta = \pi$ (0) which is independent of $\theta_1$ and $\theta_2$. Since $H_{eff}$ has zero trace, and two degenerate solutions are desired in $\varphi = \pi$ ($\varphi = 2\pi$) plane, the rest eigenvalues must satisfy $\lambda_2 = \lambda_3 = 1/2$ ($-1/2$). Consequently, the degenerate bands occur at $\beta = \pi/3$ ($\varphi = \pi$) and $\beta = 2\pi/3$ ($\varphi = 2\pi$), which fully confirm the predicted TWPs locations. Thus, our analysis reveals that while the permittivities of APTCs are modulated temporally, the refractive index of extraordinary wave propagating in a special direction along which TWPs locate remains constant.

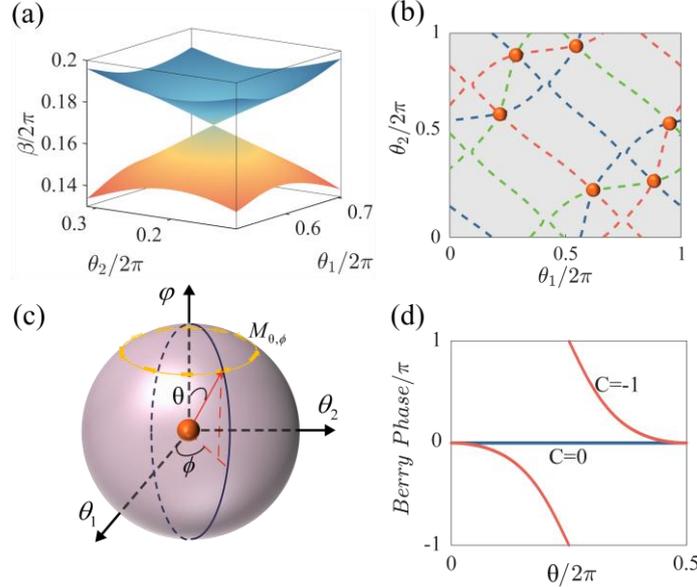

FIG. 2. (a). Band structures in the vicinity of the temporal Weyl point with $(\theta_1, \theta_2) \approx (1.241\pi, 0.454\pi)$ in $\varphi = \pi$ plane. (b). Position of TWPs in the synthetic reciprocal space. The dark blue dashed line corresponds to the condition for $n_2^{(1)} = n_2^{(2)}$, while red for $n_2^{(1)} = n_2^{(3)}$ and green for $n_2^{(2)} = n_2^{(3)}$, their intersection points are marked by orange dots, coincident with the positions of TWPs. (c). Schematic diagram of a sphere in synthetic reciprocal space. (d). Berry phase changes with polar angle $\theta$, the red curve is evaluated from the sphere wrapping a TWP, while the blue curve from the sphere without a TWP. C denotes the number of topological charges. Other parameters taken as $\varepsilon_\perp = 4.6$, $\varepsilon_P = 1.2$, $\delta\varepsilon_\perp = 4$, $\delta\varepsilon_P = 1$ and $T = 1$.

The topological charge of the TWPs can be evaluated by analyzing the left and right eigenvectors $|u_{L,n}\rangle$ and $|u_{R,n}\rangle$ of each band with the momentum eigenvalue equation. As shown in FIG. 2(c), we define a spherical manifold enclosing a TWP with $(\theta_1, \theta_2, \varphi) \approx (1.241\pi, 0.454\pi, \pi)$, and calculate Wilson loop operator $M_{\theta,\phi}$ performed along the polar angle $\phi$ at a fixed azimuthal angle $\theta$ for the lower band of TWP[50]. Berry phase obtained from Wilson loop exhibits a winding behavior as $\theta$ varies from 0 to $\pi$ [red curve in FIG. 2(d)], indicating nontrivial topological charge of the corresponding TWP. In contrast, when a sphere encloses no TWP, Berry phase [blue curve in FIG. 2(d)] remains zero, signifying trivial topology. Using this approach, topological charges of TWPs in FIG. 2(b) are determined and denoted by '+ (-)' sign for values of +1(-1).

**Fermi arcs in APTCs**

To explore the topological responses of TWPs, we construct a temporal supercell structure comprising $3p$ temporal slabs (containing $p$ unit-cells of APTC) terminated by a TCS with isotropic permittivity $\varepsilon_{TCS}$ and modulation duration $T_d = T\sqrt{\varepsilon_{TCS}}/\sum_{m=1}^{3}n_2^{(m)}$, as shown in FIG. 3(a). When the entire temporal structure undergoes periodic evolution, TCS acts as a temporal boundary after $p$ unit cells of APTC, enabling the study of finite-size effects in the time domain[24,51]. By applying the temporal transfer matrix method[50,52], we can obtain the recursive relation of the wave over time, given by $\psi_2(t) = M_2^{(t\leftarrow 0)}\psi_2(0^-)$, where $M_2^{(t\leftarrow 0)}$ describes the evolution of the extraordinary wave from the initial time to moment $t$. For an incident wave $\psi_2(0^-) = [1, \ 0]^T$, the reflection rate and transmission rate can be evaluated as $r(t) = M_{12} = [M_{21}]^*$ and $\gamma(t) = M_{11} = [M_{22}]^*$, where $M_{ij}$ are the elements of $M_2^{(t\leftarrow 0)}$. When the wave traverses the entire temporal supercell, its transfer matrix satisfies $\det\{M_2^{(T_{tot}\leftarrow 0)} - e^{j\Omega T_{tot}}I_2\} = 0$, where $T_{tot} = pT + T_d$ is the duration of temporal supercell. Given that the determinant of the transfer matrix is unity, $\det\{M_2^{(T_{tot}\leftarrow 0)}\} = 1$, $\gamma(T_{tot})$ and the corresponding quasi-frequency $\Omega$ satisfies[24]

$$\mathrm{Re}[\gamma(T_{tot})] = \cos(\Omega T_{tot}). \tag{7}$$

This relationship predicts the existence of eigenmodes with real-valued $\Omega$ when the transmission rate satisfies $\mathrm{Re}[\gamma(T_{tot})] \leq 1$, providing an efficient metric to detect Fermi arcs modes of the temporal supercell in synthetic reciprocal space. Narrow bands form and connect TWPs with opposite topological charge in synthetic reciprocal space when transmission rate satisfies $\log_{10}|\mathrm{Re}[\gamma(T_{tot})]|^2 \leq 0$, signifying the existence of Fermi arc. The evolution of the Fermi arc in the range of $\pm 0.06\pi$ around $\beta = \pi/3$ is examined in FIG. 3(b), which reveals two distinct evolutionary behaviors in opposite directions. As $\beta$ increases, the Fermi arcs contract toward the center of synthetic reciprocal space and eventually vanish, whereas decreasing $\beta$ drives them outward until they dissipate.

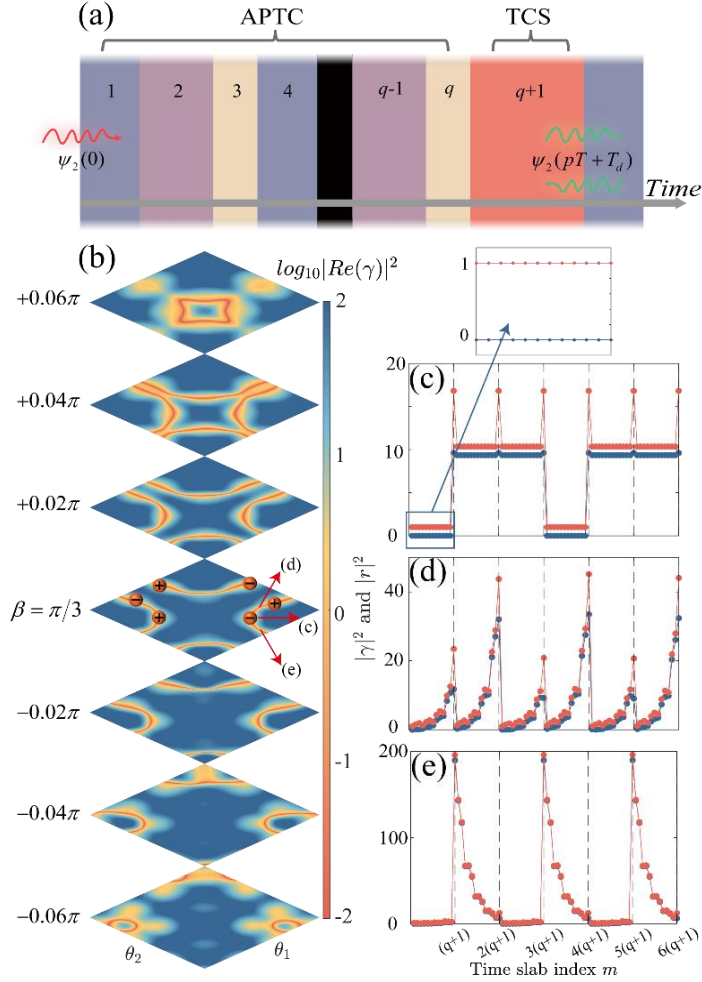

FIG. 3. (a). Schematic diagram of a temporal supercell structure assembled by $q = 12$ temporal slabs (corresponding to $p = 4$ unit cells), with TCS adding at the end of supercell structure. (b). Evolution of Fermi arcs in the supercell around $\beta = \pi/3$. The part of $\log_{10}|\text{Re}[\gamma(T_{tot})]|^2 \leq 0$ is the excitation mode in the momentum bandgap. The '+' ('−') denote a topological charge of +1 (−1). (c)-(e). Variation of transmission rate $\gamma$ (red) and reflection rate $r$ (blue) over six supercells at TWP with $(\theta_1, \theta_2) \approx (1.241\pi, 0.454\pi)$ in (c), Fermi arc with $(\theta_1, \theta_2) = (1.49\pi, 0.567\pi)$ in (d) and Fermi arc with $(\theta_1, \theta_2) = (1.133\pi, 0.2867\pi)$ in (e). The dotted lines indicate the location of 6 TCSs, where the permittivity of TCS is $\varepsilon_{TCS} = 100$.

We next calculate the variations in $\gamma(t)$ and $r(t)$ at each temporal interface via transfer matrix. FIG. 3(c) shows the evolution of $\gamma(t)$ (red curve) and $r(t)$ (dark blue curve) over time, with the positions of TCSs marked by black dashed lines. Notably, the extraordinary wave at TWPs exhibits temporal dynamics resembling those of an isotropic uniform medium: the transmission rate stays at unity ($\gamma = 1$) before the introduction of TCS and abruptly switches to a different constant afterward. This observation directly corroborates our earlier analysis of TWPs. We further examine the temporal evolution of extraordinary wave by selecting parameters along the temporal Fermi arcs. Intriguingly, the localization of these temporal Fermi arcs switches between left and right temporal boundaries after traversing a TWP. For example, excited states selected from Fermi arcs separated by one TWP display localization at right and left temporal boundaries in FIG. 3(d) and FIG. 3(e), respectively. Under time-domain excitation, the fully localized profile of the Fermi-arc mode is established only after the wave encounters the temporal boundary, a dynamic buildup constrained by causality: during propagation in the first supercell, the wave remains unaware of the forthcoming TCS and therefore cannot respond to future events. Moreover, because TWPs are topologically protected, temporal Fermi arcs that connect TWPs with opposite topological charges are robust. Here, a large $\varepsilon_{TCS} = 100$ is chosen to emulate a "hard wall" temporal boundary. However, the existence of the Fermi arc is topologically guaranteed and persists for a wide range of $\varepsilon_{TCS}$ values[28,29,50].

**Radiation control of a stationary charge**

Given a stationary charge $Q$ embedded in an APTC can radiate electromagnetic wave[26], where discontinuity in electric displacement vector $\vec{D}_0^m = \frac{jQ}{k}\frac{1}{\varepsilon_\perp^{(m)}(\hat{k}_x^2+\hat{k}_y^2)+\varepsilon_P^{(m)}\hat{k}_z^2}\left[\varepsilon_\perp^{(m)}\hat{k}_x,\ \varepsilon_\perp^{(m)}\hat{k}_y,\ \varepsilon_P^{(m)}\hat{k}_z\right]^T$ across temporal interfaces acts as the effective source that generate the radiative fields, we investigate how temporal Weyl points (TWPs) and their associated Fermi arcs influence this radiation process. Due to the continuity of the radiation fields in time domain, their behavior across a temporal interface $t_s$ is governed by the matching relation[50]:

$$\psi_2(t_s^+) = s_2^{(m+1,m)} + J_2^{(m+1,m)}\psi_2(t_s^-), \tag{8}$$

where $s_2^{(m+1,m)} = \left[h^{(m)} - h^{(m+1)},\ 0\right]^T$ is the extra source term produced at the interface between slab $m$ and $m$+1 with $h^{(m)} = \frac{jQ}{k}\cdot\frac{\cos\theta\left[\varepsilon_\perp^{(m)} - \varepsilon_P^{(m)}\right]}{\varepsilon_\perp^{(m)}\sin^2\theta + \varepsilon_P^{(m)}\cos^2\theta}$. Assuming no radiation field exists prior to $t = 0$ [i.e., the initial field $\psi_2(0^-) = 0$], the radiation field after $p$ cycles is derived from Eq. (8) as:

$$\psi_2(pT^-) = \left[J_2^{(1,3)}\right]^{-1}\cdot\left[\frac{I_2 - M_2^p}{I_2 - M_2}N - s_2^{(1,3)}\right], \tag{9}$$

where $M_2 = J_2^{(1,3)}F_2^{(3)}J_2^{(3,2)}F_2^{(2)}J_2^{(2,1)}F_2^{(1)}$ is the transfer matrix for a unit-cell APTC and $N = s_2^{(1,3)} + J_2^{(1,3)}F_2^{(3)}s_2^{(3,2)} + J_2^{(1,3)}F_2^{(3)}J_2^{(3,2)}F_2^{(2)}s_2^{(2,1)}$. We calculate the radiation energy density $w_{EM}^{(r)}(\vec{k},t) = 2\sin^2\theta|\psi_2(t)|^2/\varepsilon_0\left[n_2^{(m)}\right]^2$ at the final slab of APTC with a realistic charge value $Q = 10^{-6}C$. Whereas the radiation energy from stationary charges within band structures of an APTC remains bounded in time, it grows exponentially when the excitation parameters are selected within the momentum bandgap, where $\text{Im}(\Omega) \neq 0$ [26].

To investigate the influence of TWPs on the radiation, we choose specific parameters $(\theta_1,\theta_2) \approx (1.241\pi, 0.454\pi)$ in synthetic reciprocal space, where TWPs exist along $\theta = \pi/6$ and are marked by green-colored dots in FIG. 4(a). After $p$=60 modulation periods, the radiation energy density grows exponentially in all directions except along $\theta = \pi/6$ [denoted by the red line in FIG. 4 (a)], indicating the presence of "near-zero radiation" in that direction, compared to the radiation in momentum bandgaps (see the rigorous proof for the existence of eigenstates along $\theta = \pi/6$ in ref.[50]). TWPs carrying opposite topological charges exhibit no radiation growth over time. Crucially, they are characterized by distinct radiation signatures: one is located at a radiation peak, and the other adjacent to a valley [FIG. 4(e)], confirming that these features are signatures of the topological charge rather than simple stability boundaries. In contrast, we select a point in the synthetic reciprocal space where TWPs are absent at $(\theta_1,\theta_2) \approx (1.49\pi, 0.566\pi)$, and recalculate the radiation energy density. As shown in FIG. 4(b), "near-zero radiation trajectory" disappears, as the momentum bandgaps now extend across all direction. Interestingly, near TWPs, the frequencies and directions of dominant radiation vary rapidly. These intriguing properties enable abrupt changes in the principal radiation direction with slight parameter tuning, which is analyzed in details in [50].

To study the effects of Fermi arcs on the radiation from stationary charges, we introduce TCSs to create temporal boundaries of APTCs and reevaluate the resulting radiation energy density $w_{EM}^{(r)}(\vec{k},T_{tot})$ using Eq. (8). Since these temporal boundaries give rise to topological temporal Fermi arcs, the associated radiation energy transitions from exponential growth to a confined, steady level over time. For example, when we chose $(\theta_1,\theta_2) = (1.49\pi, 0.566\pi)$, the radiation energy density along $\theta = \pi/6$ direction at positions $k = (3l+1)k_0\sum_{m=1}^{3}n_2^{(m)}/6$ [with $l = 0,1,2,...$ and $k_0 \equiv 2\pi/Tc_0$] and $k = (3l+2)k_0\sum_{m=1}^{3}n_2^{(m)}/6$, corresponding to momentum phase $\beta = \pi/3 + l\pi$ and $\beta = 2\pi/3 + l\pi$, grows exponentially over time [marked by blue dots in FIG. 4(b)]. However, after embedding a TCS every 5 or 6 cycles (15 or 18 slabs) in the APTCs, these positions exhibit nearly zero radiation compared to other regions, as shown in FIGs. 4(c) and 4(d), since

Fermi arcs appear and occupy these locations due to the presence of temporal boundaries. Meanwhile, the supercell structure introduces numerous new bands and associated momentum bandgap, altering the radiation-energy distribution. Although the positions of these bands shift in *k*-space as the number of unit cells in a supercell varies, the temporal Fermi arcs persist and retain their *k*-space locations due to the topological protection of TWPs[50]. Consequently, the low-radiation regions associated with Fermi arcs remain fixed, whereas area without Fermi arcs ($\beta = 2\pi/5$) marked by red dot in FIG. 4(c) exhibits increased radiation in FIG. 4(d) after changes in the supercell.

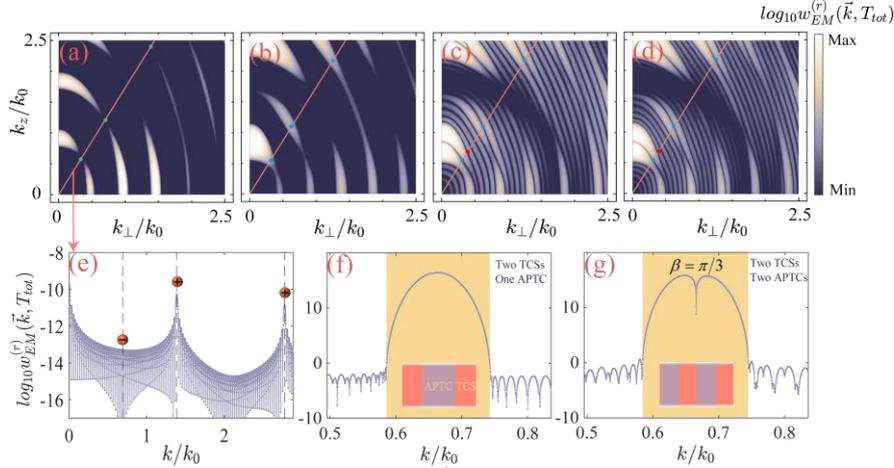

FIG. 4. Radiation energy density in logarithm scale $\log_{10} w_{EM}^{(r)}(\vec{k}, T_{tot})$ for an APTC with an embedded stationary charge after 60 cycles (*p*=60). (a) Radiation distribution at position $(\theta_1, \theta_2) = (1.241\pi, 0.454\pi)$ in synthetic reciprocal space, which contains TWPs along $\theta = \pi/6$. The pink line denotes the direction of $\theta = \pi/6$, and the green dots mark the momentum phase $\beta = \pi/3 + l\pi$ [$k = (3l+1)k_0 \sum_{m=1}^{3} n_2^{(m)}/6$] and $\beta = 2\pi/3 + l\pi$ [$k = (3l+2)k_0 \sum_{m=1}^{3} n_2^{(m)}/6$]. (b) Radiation distribution at position $(\theta_1, \theta_2) = (1.49\pi, 0.566\pi)$ in synthetic momentum space. (c), (d) Radiation distribution for the temporal supercells, in which one TCS is added every (c) five, (d) six cycles. The red dot marks the momentum phase $\beta = 2\pi/5$ [$k = k_0 \sum_{m=1}^{3} n_2^{(m)}/5$]. (e) The variations of "near-zero radiation" in (a) with the dotted lines indicate the location of TWPs. (f-g) The radiation variations along the $\theta = \pi/6$ near the momentum phase $\beta = \pi/3$ [$k = k_0 \sum_{m=1}^{3} n_2^{(m)}/6$], consider another position of the synthetic momentum space $(\theta_1, \theta_2) = (1.136\pi, 0.2867\pi)$. Configuration (f) has two TCSs inserted at the beginning and end, while (g) has one TCS inserted every 30 cycles, for a total of two TCSs. The momentum bandgap is denoted by yellow-shaded region.

Finally, we investigate the effects of TCS on the radiation of stationary charges associated with temporal Fermi arcs. We consider a 60-cycle supercell with either two TCSs at the beginning and end of the supercell [in FIG. 4(f)] or two TCSs every 30 cycles [in FIG. 4(g)], and calculate the radiation across the *k*-spectra spanning the momentum bandgap (yellow-shaded region) along $\theta = \pi/6$. The stationary charge emits strongly throughout whole momentum bandgap in FIG. 4(f), whereas low radiation appears at frequency $k = k_0 \sum_{m=1}^{3} n_2^{(m)}/6$ ($\beta = \pi/3$) in FIG. 4(g). These results indicate the radiation suppression is not simply due to the presence of TCSs, but arises from the formation of topological Fermi arc modes, which requires wave to travel at least two supercells during the temporal evolution, consistent with the results in FIG. 3. Thus, the emergence of temporal Fermi arcs induced by TCS insertion selectively suppresses radiation bursts at particular directions and frequencies, offering a robust mechanism to control the radiation of stationary charges.

In conclusion, our study establishes APTCs as a powerful synthetic platform for realizing Weyl physics in the time domain. By constructing topological TWPs in synthetic reciprocal space and uncovering their associated Fermi arc states at temporal boundaries, we demonstrate that time-varying systems can host robust topological phenomena previously thought to be confined to spatially periodic structures. Importantly, we reveal that these topological features profoundly alter the radiation behavior of stationary charges, with TWPs enabling tunable directional radiation and Fermi arcs suppressing emission at selected direction and frequencies. These findings not only enrich the fundamental understanding of time-modulated media but also open new routes for topological control of light–matter interactions, with promising implications for broadband radiation engineering, non-Hermitian photonics, and synthetic-dimensional and time-varying quantum platforms.

the details of topological charge caculation, the edge states of temporal supercell, several derivations of stationary charge radiation in APTC and discussion of dominant radiation control.